# Layer-Selective Magnetization Switching in the Chirped Photonic Crystal with GdFeCo

O.V. Borovkova*[1,2], D.O. Ignatyeva[1,2], and V.I. Belotelov[1,2]

[1]*Faculty of Physics, Lomonosov Moscow State University, Moscow, Russia*
[2]*Russian Quantum center, Skolkovo, Russia*

*Corresponding author e-mail: o.borovkova@rqc.ru*

**Abstract.** Here we propose a magnetophotonic structure for the layer-selective magnetization switching with ultrashort laser pulses of different wavelengths. It is based on a chirped magnetophotonic crystal (MPC) containing magnetic GdFeCo and nonmagnetic dielectric layers. At each operating wavelength the laser pulses heat up to necessary level only one GdFeCo layer that leads to its magnetization reversal without any impact on the magnetization of the other layers. Moreover, magneto-optical reading of the MPC magnetization state is discussed. Lateral dimensions of the considered MPC can be made small enough to operate as a unit cell for data storage.

## I. Introduction

Up-to-date technologies require modern information devices to process large amounts of data at high rates. Methods of the ultrafast magnetism are very promising in this respect since they allow controlling the magnetization of a magnetic material by the femtosecond laser pulses at a picosecond time scales [1-10].

Ferrimagnetic alloy of gadolinium-iron-cobalt (GdFeCo) is one of the most promising materials for further development of the information storage devices, since it allows to perform a single-shot all-optical magnetization switching during a few picoseconds [11-13]. in the area of pulse action on the GdFeCo film. A mechanism of all-optical switching originates from the ultrafast spin dynamics of the Gd and FeCo sublattices which are thrown out of balance due to the light-induced thermal heating [14, 15]. Thus, ultrafast all-optical magnetization switching depends on the amount of energy delivered by the optical pulse [16] and is characterized by a threshold femtosecond pulse intensity required for the magnetization switching [17, 18].

Nowadays many efforts are put to study how to control the process of all-optical switching via tailoring the properties of the femtosecond pulse [18-20] and via the additional patterning of the magnetic structure and surface plasmon excitation [21-24]. The latter is responsible for strong light localization [23-26] which enables layer-selective all-optical switching in a bilayer heterostructure of GdFeCo that has been demonstrated recently [23]. However, the disadvantage of the SPP-assisted selectivity [23] is that it intrinsically requires large amounts of metal to support SPP, and, consequently, this approach could not be extended to multilayered structures with large number of GdFeCo layers in a multilayer stack.

In its turn, the electromagnetic field distribution can be reconfigured by the frequency detuning in a specially designed multilayer nanostructures with certain optical properties [27, 28]. The magnetophotonic crystal (MPC) structures [29-32], i.e. the multilayer nanostructures containing the magnetic layers were shown to perform tunable localization of light at the resonant frequencies enabling the enhancement of the optomagnetic interaction in a magnetic material [32].

Here we propose a magnetophotonic crystal with thin smooth or perforated layers of GdFeCo ferrimagnetic alloy that provides the local magnetization control and targeted remagnetization of the single layers of the multilayer structure. By the frequency tuning of the input femtosecond laser pulses one can select the magnetic layer of GdFeCo in which the energy of the laser pulse is mostly concentrated and exceeds the magnetization reversal threshold, while in others it is 1.5-2 times lower. In other words, the proposed MPC allows to 'turn on' one certain magnetic layer of the structure and leave all other layers 'turned off'. It is shown that in the addressed MPC structure the magnetic layer of GdFeCo can be perforated to separate bits from each other in lateral direction. We also provide the method of an all-optical single-wavelength information reading in this multilayered structure. Thus, the designed MPC with GdFeCo layers could serve as a basis of 3D information storage device.

## II.     A Design of the Chirped Magnetophotonic Crystals with GdFeCo Layers

Let's consider a 1D chirped MPC structure composed of the four pairs of non-magnetic layers of titanium dioxide ($TiO_2$) and silicon dioxide ($SiO_2$) with the thicknesses gradually increasing deeper into the MPC (see Fig. 1). The thicknesses of the $TiO_2$ and $SiO_2$ layers are given in the Table 1.

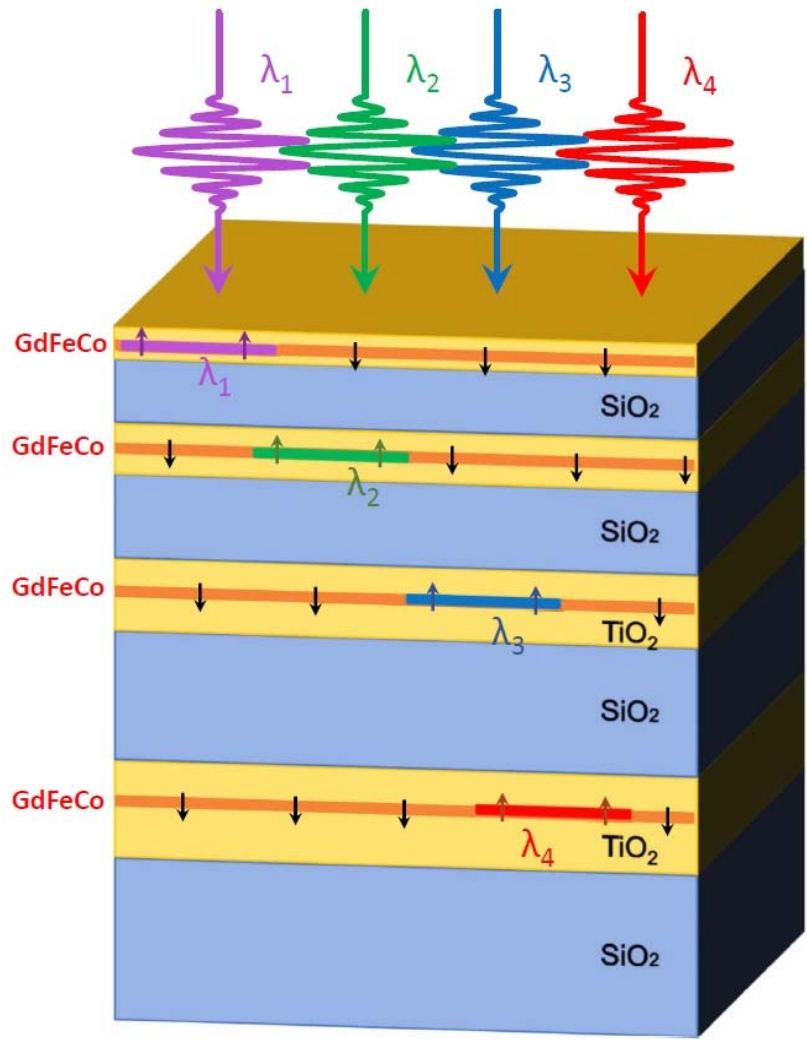

Figure 1. A scheme of the addressed chirped magnetophotonic crystal with thin GdFeCo layers. $TiO_2$ layers (yellow) alternate with $SiO_2$ layers (blue). Orange areas indicate the GdFeCo layers introduced into $TiO_2$. Depending on the frequency of the input femtosecond pulse the different GdFeCo layers experience remagnetization due to the heating. This process is illustrated by multi-colored rectangles. The magnetization direction in GdFeCo layers is shown by small arrows. The dimensions and proportions are not respected and are chosen just for ease of perception (for the correct dimensions see Table 1).

|  | Pair 1 | | Pair 2 | | Pair 3 | | Pair 4 | |
|---|---|---|---|---|---|---|---|---|
| Material | TiO$_2$ | SiO$_2$ | TiO$_2$ | SiO$_2$ | TiO$_2$ | SiO$_2$ | TiO$_2$ | SiO$_2$ |
| Thickness (nm) | 48 | 270 | 61 | 292 | 75 | 317 | 83 | 330 |

Table 1. The layer thicknesses of the addressed chirped magnetophotonic crystal.

The variation of the layer's thickness deeper inside the chirped MPC structure leads to the shift of the bandgap corresponding to each TiO$_2$/SiO$_2$ pair. Varying the frequency of the input pulses on may achieve that the light will reach only a predefined pair of layers, but the deeper layers will remain practically inaccessible to the pulses because of the bandgap.

The energy of the EM field is mostly concentrated in the layers with higher refractive index at the long-wavelength edge of the bandgap (and vice versa). As soon as we operate at the long-wavelength edge of the bandgap due to the growth of the layer's thickness deeper inside the MPC, we expect higher concentration of the EM field in the TiO$_2$ layers having greater refractive index. Actually, the refractive index of SiO$_2$ is $n_{SiO_2} = 1.45$, and for TiO$_2$ layers $n_{TiO_2} = 2.58$ at $\lambda = 644 nm$ and they both increase for longer wavelengths [33]. Therefore, we introduce the thin layers of GdFeCo into the TiO$_2$ layers to provide the highest concentration of the EM field in them. The thin magnetic layers of GdFeCo are introduced inside the TiO$_2$ layers so that the total thickness of the complex TiO$_2$ and GdFeCo layer remains as it is given in the Table 1. Substituting some parts of TiO$_2$ by GdFeCo ultrathin layers makes just minor changes of the optical properties of the MPC, but at the same time it ensures a strong localization of light in the layers with magneto-optical response at the 'working' wavelengths. We address two types of the GdFeCo layer. The first one is a smooth layer with the thickness of 3 nm. The second type is the perforated 9nm-thick layer. The perforation has a form of the 1D subwavelength grating with a period of 90 nm. 30nm-thick stripes of GdFeCo alternate with 60nm-thick stripes of TiO$_2$. Such perforated layer type where magnetic cells are separated from each other both in depth as well as in the lateral direction provides the additional opportunities for the 3D data storage devices.

The thin layers of GdFeCo can serve as cells for the bit data storage. SiO$_2$ layers with the thickness of about 300 nm act as a buffer layers and provide a possibility to switch the magnetization in the single data cell not affecting the adjacent cells.

### III. Selective Remagnetization of the GdFeCo layers in the Chirped Magnetophotonic Crystal

If the MPC is illuminated by ultrashort pulses falling normally to the MPC layers then spatial distribution of the EM field in the structure depends on the input light frequency. For the addressed MPC structure, the spectral distribution of the averaged square of the electric field absolute value, <|E|$^2$>, in the TiO$_2$ layers is given in Fig. 2. For each of the four layers one can choose the wavelength when the EM energy in that layer exceeds significantly the energy in the other layers. Particularly, at the wavelengths of 644 nm, 686 nm, 832 nm, and 994 nm the light is mostly localized inside the first, second, third and fourth TiO$_2$ layer of the proposed MPC, correspondingly.

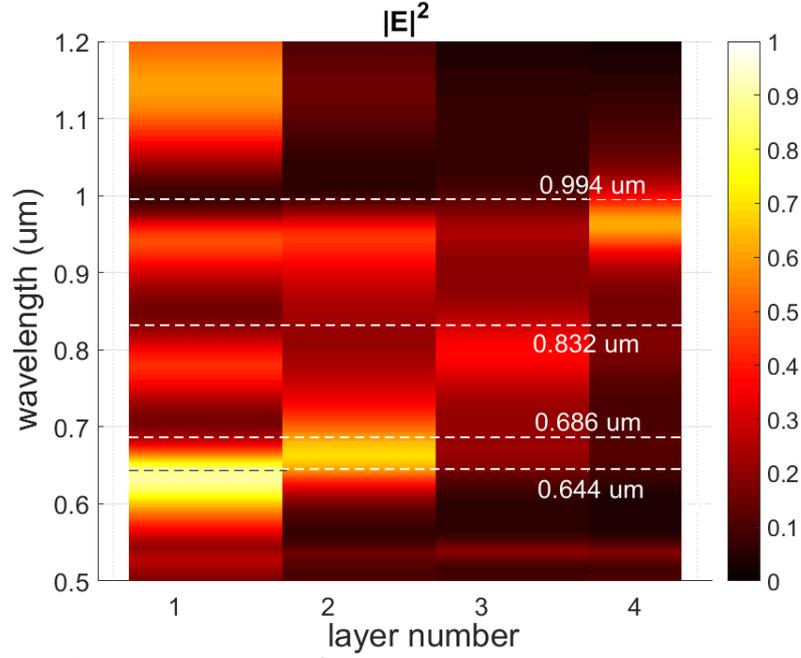

Figure 2. A spectral distribution of the mean |E|² in TiO₂ layers with the smooth 3nm-thick GdFeCo layers inside. Dashed lines indicate the operating wavelengths.

As it was mentioned above the femtosecond laser pulses propagating through the magnetic material heat the layers. The heating of a certain layer is determined by the energy of the laser pulse, i.e. |E|², in that layer. The spatial distribution of |E|² in the chirped MPC at the chosen operating wavelengths is given in Fig. 3. The value of |E|² is given normalized to its maximum value at the certain wavelength. The spatial distributions of |E|² as well as the spectral distribution in Fig. 2 have been numerically calculated by means of the rigorous coupled-wave analysis (RCWA) [34, 35]. The choice of the method is due to the fact that it is suitable for the numerical analysis of both types of the GdFeCo layers addressed here, smooth and perforated layers.

The yellow areas in Fig. 3 correspond to TiO2 layers, the blue regions refer to SiO2 layers, and the red lines indicate the position of the 3nm-thick GdFeCo layers. The positions of GdFeCo layers are chosen to coincide with the maximum value of |E|² at each of four operating wavelengths, that provides the highest possible energy concentration inside the magnetic layer and its proper heating as well.

The first GdFeCo layer is placed 19 nm in depth of the upper TiO₂ layer and gets highest optical energy at λ=644 nm (Fig. 3a). At that wavelength the EM field in the first GdFeCo layer ($\left|E\right|_1^2$) exceeds the field in the other layers. In particular, the EM field in the second GdFeCo layer is $\left|E\right|_2^2 = 0.63\left|E\right|_1^2$, and the EM field in the third and fourth GdFeCo layers do not exceed 10% of $\left|E\right|_1^2$. Therefore, operating at the wavelength of 644 nm one can guarantee that light is mostly localized in the first magnetic layer and the magnetization switching takes place only in this layer. As a result, the first GdFeCo layer becomes 'turned on' while the rest magnetic layers remain untouched.

On the other hand, the second GdFeCo layer is put 40 nm inside the second TiO₂ layer which ensures maximum concentration of the optical energy in this layer at λ=686 nm. The third and fourth layers of GdFeCo are buried inside the corresponding TiO₂ layer 21 nm and 83 nm in depth, respectively.

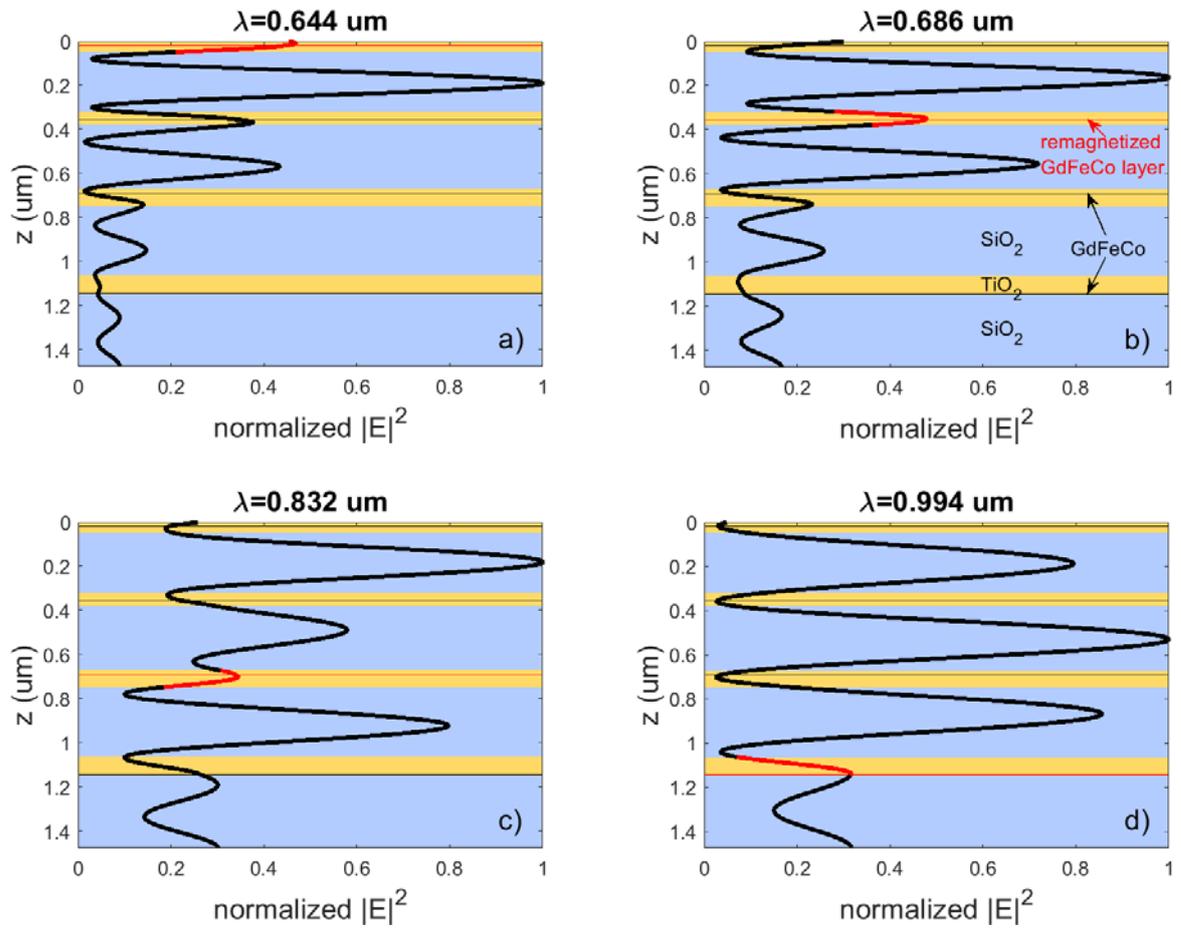

Figure 3. The spatial distribution of the normalized |E|² in the chirped MPC at the 'operating' wavelengths of (a) 644 nm, (b) 686 nm, (c) 832 nm, (d) 994 nm. Red sections of the curve show the magnitude of the normalized |E|² inside the addressed GdFeCo layer and nearby. The yellow areas correspond to $TiO_2$ layers, the blue regions refer to $SiO_2$ layers, and the red and black lines indicate the position of the 3nm-thick GdFeCo layers. The red ones denote the position of the remagnetized layers, and the black ones refer to the GdFeCo layers with untouched magnetization.

Table 2. The relative values of the maximum of |E|² in the magnetic layers at the operating wavelengths in case of the smooth 3nm-thick GdFeCo layer.

| Wavelength | Max|E|² in 1st GdFeCo layer | Max|E|² in 2nd GdFeCo layer | Max|E|² in 3d GdFeCo layer | Max|E|² in 4th GdFeCo layer |
|---|---|---|---|---|
| λ=0.644um | 1 | 0.6286 | 0.0710 | 0.0866 |
| λ=0.686um | 0.3519 | 1 | 0.2011 | 0.1812 |
| λ=0.832um | 0.4040 | 0.5605 | 1 | 0.7235 |
| λ=0.994um | 0.0982 | 0.0896 | 0.0824 | 1 |

Table 3. The relative values of the maximum of |E|² in the magnetic layers at the operating wavelengths in the case of the perforated 9nm-thick GdFeCo layer.

| Wavelength | Max|E|² in 1st GdFeCo layer | Max|E|² in 2nd GdFeCo layer | Max|E|² in 3d GdFeCo layer | Max|E|² in 4th GdFeCo layer |
|---|---|---|---|---|
| λ=0.644um | 1 | 0.6530 | 0.0894 | 0.0916 |
| λ=0.686um | 0.4686 | 1 | 0.2286 | 0.1831 |
| λ=0.832um | 0.3274 | 0.3572 | 1 | 0.6932 |
| λ=0.994um | 0.1642 | 0.1596 | 0.1497 | 1 |

Tables 2 and 3 give the detailed information on the ratio of the field values in different magnetic layers at the operating wavelengths for two addressed configurations. Table 2 refers to the MPC with smooth 3nm-thick GdFeCo layers, and Table 3 presents data on the MPC configuration with 9nm-thick perforated GdFeCo layers.

The data in the Tables 2 and 3 is given in the relative units. It means that the mean EM field $|E|^2$ in the selected layer for each of the operating wavelengths is taken as 1 and the mean EM field in all others layers is compared with the field in the selected layer. It allows to show clearly that both types of the proposed chirped MPC allow to obtain a necessary excess of the EM field in the chosen layer compared to other magnetic layers. That excess is not less than 30%. That dramatic difference of the EM field energy in different magnetic layers of the MPC makes it possible to provide rather different levels of heating in different GdFeCo layers. Consequently, the considered chirped MPC structure allows to record information into each of the layers independently by femtosecond pulses at one of four wavelengths.

### IV. Magnetization Probing of the Chirped Magnetophotonic Crystal

Now we discuss a method to probe the magnetization state of the magnetic layers in the proposed chirped magnetophotonic structure. There are various complicated methods for depth-resolved magnetization probing, for instance, by means of x-rays [36] etc. However, the method proposed here is based on the measurement of the Faraday effect [37-39] at a single wavelength of laser diode and could be easily implemented in data reading devices. The chosen parameters of the structure provide not only the heating above the threshold in different layers at the different operating wavelengths, but also different Faraday rotation angles by the magnetic layers. By measuring the magnitude of the Faraday effect at a certain wavelength, it is possible to determine the magnetization of each of the layers of the chirped MPC. The similar approach has been proposed earlier in Ref. [40] for the two-layer structure.

We determine the sensitivity to the magnetization of the j-th GdFeCo layer as follows

$$S_j = \frac{|\Phi_j|}{\sum_{k=1}^{N}|\Phi_k|},$$

where $\Phi_j$ is a Faraday rotation angle of the linearly polarized light coming through the chirped MPC with magnetized j-th GdFeCo layer, $N$ is a total number of the GdFeCo layers in the chirped MPC (here $N=4$).

In the Fig. 4 the layer sensitivity $S_j \left( j = \overline{1,4} \right)$ versus the probing light wavelength is shown. The Faraday effect of the chirped MPC depends strongly on the index number of the magnetized layer. Measuring the spectrum of the Faraday effect of the structure one can determine which GdFeCo layers are magnetized. Thus, one can read the information from such 3D nanostructure.

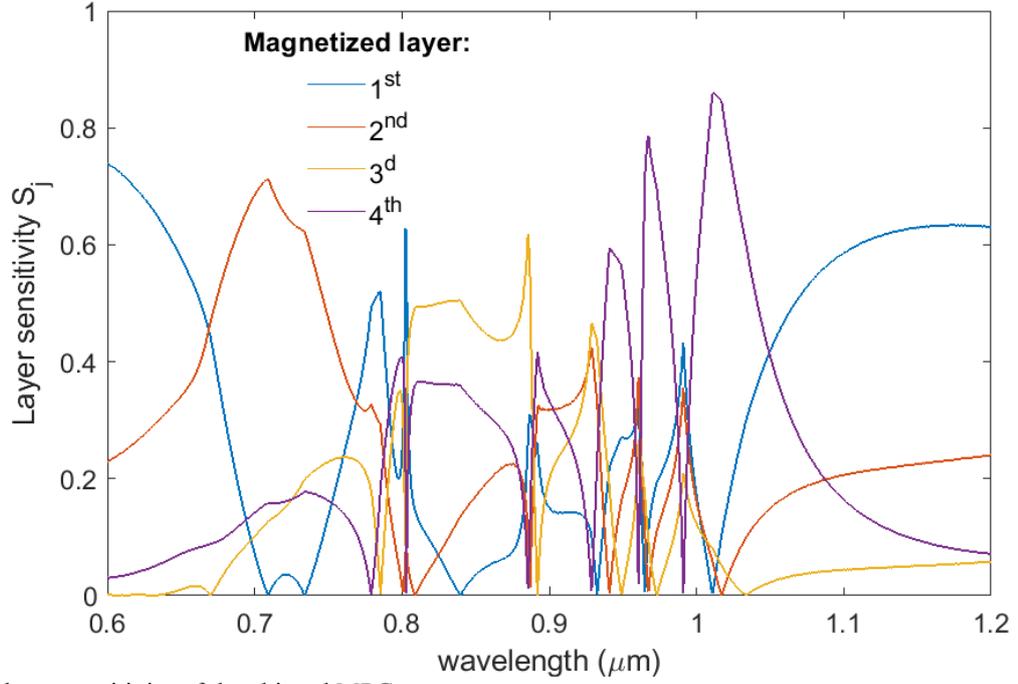

Figure 4. A layer sensitivity of the chirped MPC.

Besides that, operating at the certain wavelength one can probe the magnetization of all four magnetic layers of the structure. The operating wavelength should satisfy the following conditions. Firstly, the rotation angle of the polarization plane is noticeably different for each of the magnetic layers. Secondly, no combination of sums of Faraday angles from different magnetized layers is equal to another Faraday angle or other combination of sums. In other words, the Faraday rotation from one or several magnetized GdFeCo layers is different from the Faraday rotation provided by other magnetized layers of the chirped MPC. This method allows us to determine which of the magnetic layers are magnetized, that is, it will allow reading the information recorded in the structure.

The most convenient, of course, is to operate at the wavelengths of the laser diodes. For example, the Nd laser diodes with the output wavelength of 1.064um are widespread and available with various output powers. In the Table 4 the layer sensitivity $S_j \left( j = \overline{1,4} \right)$ are given for three wavelengths corresponding to the output wavelength of the industrial laser diodes, 0.980um, 1.064um, and 1.083um.

Table 4. The layer sensitivity at three certain wavelengths, 0.98um, 1.064um, and 1.083um.

| Wavelength | $S_1$ | $S_2$ | $S_3$ | $S_4$ |
|---|---|---|---|---|
| λ=0.980um | 0.2344 | 0.1673 | 0.0593 | 0.5390 |
| λ=1.064um | 0.4888 | 0.1750 | 0.0332 | 0.3030 |
| λ=1.083um | 0.5512 | 0.1947 | 0.0405 | 0.2136 |

In order to unambiguously probe the magnetization of a particular magnetic layer of the structure, it is necessary to ensure that two conditions are met. First, the Faraday rotation experienced by the probing light is different in case of magnetization of different GdFeCo layers. Second, no combination of the sums of values of Faraday rotations in different layers gives a magnitude of Faraday rotation of the plane of polarization in other layers. Analyzing the data given in Table 4 one can make sure that the first condition is met, since for any wavelength the sensitivity to magnetization of different layers is noticeably different.

Then, in order to make sure that the second condition is satisfied, we will perform the following procedure. First of all, we sort the original sets of $S_j \left(j=\overline{1,4}\right)$ so that the resulting set $f_j \left(j=\overline{1,4}\right)$ is the ascending sequence ($f_1 < f_2 < f_3 < f_4$). Further, we compose various combinations of the sums $f_1+f_2$, $f_2+f_3$ etc. and compare them with the other elements of the sequences $f_j \left(j=\overline{1,4}\right)$. For instance, the ratio $(f_1+f_2-f_3)/f_3$ shows how the sum $f_1+f_2$ differs from the value $f_3$. In particular, at the wavelength of 0.980um the sum $f_1+f_2$ differs from $f_3$ by 3.3%. Similar ratios of various combinations of variables $f_j \left(j=\overline{1,4}\right)$ are given in Table 5.

Table 5. The relative difference of the layer sensitivity at three certain wavelengths, 0.98um, 1.064um, and 1.083um.

| | λ=0.980um | λ=1.064um | λ=1.083um |
|---|---|---|---|
| $\dfrac{f_1+f_2-f_3}{f_3}$ | 3.3% | 31.3% | 10.1% |
| $\dfrac{f_1+f_2-f_4}{f_4}$ | 58% | 57.4% | 57.3% |
| $\dfrac{f_2+f_3-f_4}{f_4}$ | 25.5% | 2.2% | 26% |
| $\dfrac{f_1+f_3-f_4}{f_4}$ | 45.5% | 31.2% | 54% |
| $\dfrac{f_1+f_2+f_3-f_4}{f_4}$ | 14.5% | 4.6% | 18.6% |
| $\dfrac{f_1+f_4-f_2-f_3}{f_2+f_3}$ | 49% | 9.2% | 45% |

From analysis of the Table 5 one can made sure that the layer sensitivity values have a difference between themselves by at least 2% at the wavelengths of 0.980um, 1.064um, and 1.083um. This makes it possible to check that no combination of the Faraday rotation in some layers will be equal to the combination of the Faraday rotations in the other magnetic layers. Consequently, the proposed method of magnetization probing gives unambiguous results in the designed chirped MPC structure at any of these frequencies.

## V. Conclusions

To sum up, a design of the MPC structure providing a selective magnetization switching in the different magnetic layers at different frequencies is proposed. The MPC contains the thin layers of GdFeCo as magnetic counterparts. By illuminating the MPC by femtosecond laser pulses at the certain frequencies we create the conditions for the highest concentration of the electromagnetic field in the necessary layer of the structure while in the other GdFeCo layers the intensity is at least 1.5 times smaller. Thus, one can achieve selective all-optical magnetization reversal in a single layer of the multilayered stack determined by the laser frequency without any impact on the other layers. The magnetic layers of the MPC are separated from each other by the non-magnetic layers, and can be perforated, so, they can serve as the unit cells for information storage. We also provide a mechanism of single-wavelength reading of the information stored in such a multilayered stack. The approach was demonstrated for the 4 layers of GdFeCo, however it could be extended to the larger number of layers.

The reported study was funded by RFBR according to the research project no. 18-32-20225 mol_a_ved.